\documentclass[preprint,12pt]{elsarticle}

\usepackage{algorithm}
\usepackage{algorithmic}

\usepackage{amssymb,amsmath}
\usepackage{amsthm}
\newtheorem{definition}{Definition}[section]

\newcommand{\SP}{\textsuperscript}

\newcounter{bla}

\journal{Computer Physics Communications}

\usepackage{filecontents}

\begin{document}

\begin{frontmatter}



\title{\textit{differint}: A Python Package for Numerical Fractional Calculus}


\author[a]{Matthew Adams\corref{author}}
%\author[a,b]{Second Author}
%\author[b]{Third Author}

\cortext[author] {Corresponding author.\\\textit{E-mail address:} Matthew.Adams@ucalgary.ca}
\address[a]{Department of Mathematics and Statistics, 612 Campus Place N.W., Calgary, AB, CANADA, T2N 1N4}
%\address[b]{Second Address}

\begin{abstract}
% Acceptable program descriptions can take different forms. The following Long Write-Up structure is a suggested structure but it is not obligatory. Actual structure will depend on the length of the program, the extent to which the algorithms or software have already been described in literature, and the detail provided in the user manual.

% Your manuscript and figure sources should be submitted through the Elsevier Editorial System (EES) by using the online submission tool at \\
% http://www.ees.elsevier.com/cpc.

% In addition to the manuscript you must supply: the program source code; job control scripts, where applicable; a README file giving the names and a brief description of all the files that make up the package and clear instructions on the installation and execution of the program; sample input and output data for at least one comprehensive test run; and, where appropriate, a user manual. These should be sent, via email as a compressed archive file, to the CPC Program Librarian at cpc@qub.ac.uk.

Fractional calculus has become widely studied and applied to physical problems in recent years [2,3]. As a result, many methods for the numerical computation of fractional derivatives and integrals have been defined. However, these algorithms are often programmed in an ad hoc manner, requiring researchers to implement and debug their own code.

This work introduces the \textit{differint} software package, which offers a single repository for multiple numerical algorithms for the computation of fractional derivatives and integrals. This package is coded in the open-source Python programming language [1]. The Gr\"unwald-Letnikov, improved Gr\"unwald-Letnikov, and Riemann-Liouville algorithms from the fractional calculus are included in this package. The algorithms presented are computed from their descriptions found in [2]. This work concludes with suggestions for the application of the \textit{differint} software package.

\end{abstract}


\end{frontmatter}

\section{Introduction}
\label{sec:intro}

Fractional calculus is a generalization of the differential and integral calculus. This field of study has seen a recent rise in popularity do to its applications for the solution of fractional ordinary differential equations and fractional diffusive equations \cite{baleanu2010fractional}. Several definitions of fractional derivatives and integrals exist, including the Gr\"unwald-Letnikov, the `improved' Gr\"unwald-Letnikov \cite{oldham1974fractional}, the Riemann-Liouville, and the Caputo. For the purposes of this work, the fractional derivative and integral will be combined into one term using the nomenclature of Oldham and Spanier \cite{oldham1974fractional}, and will henceforth be referred to as `differintegrals'. This term unites the concepts of differentiation and integration in very much the same way as the fractional calculus itself.

Many algorithms have been proposed for the numerical evaluation of the various definitions of the differintegral \cite{baleanu2010fractional,oldham1974fractional,mathieu2003fractional,diethelm1997algorithm,li2016higherorder}. In this paper, the \textit{differint} package is presented. All code for this package is written in the Python programming language \cite{Rossum:1995:PRM:869369}. Python was chosen as the language of implementation due to its ease of use, wealth of available support, and the NumPy package, which makes use of BLAS level 3 algorithms for numerical linear algebra \cite{walt2011numpy}. The \textit{differint} package contains three algorithms for the respective numerical computation of the Gr\"unwald-Letnikov (GL) backward finite difference method, the improved GL centered difference method, and the Riemann-Liouville (RL) quadrature method. Section 2 introduces each algorithm and its computational complexity. In section 3, the core and auxiliary functions of the \textit{differint} package are described, with some examples of the \textit{differint} package in action. Finally, section 4 concludes with a discussion of the package and suggestions for its application.



%% References

\section{Fractional Differentiation and Integration}
\label{sec:algorithmdescription}

\subsection{The GL Algorithm}
Our first finite difference algorithm for calculating the GL differintegral was presented in \cite{oldham1974fractional} and refers directly to the following definition.

\begin{definition}\label{def:GL}
	Let $f(x)$ be an analytic function, bounded on the domain $[0,x]$, and divide the domain into $N+1$ equally spaced grid points with step size $h$. The Gr\"unwald-Letnikov differintegral of order $\alpha \in \mathbb{R}$ on the domain $[0,x]$ is defined as:
  \begin{equation}
      D^\alpha f(x) = \lim_{h\to 0} h^{-\alpha} \sum_{k=0}^{N-1} b_k f(x-kh),
  \end{equation}
  where the coefficients $b_k$ are given as $\frac{a(a+1)\cdots(a+k-1)}{k!} = \frac{(a)_k}{k!}$. 
\end{definition}

Note that we have used the Pochhammer symbol 
$$ (a)_0 = 1, \quad (a)_k = (a)\cdot(a+1)\cdots(a+k-1) $$
in Definition \ref{def:GL} (see \cite{seaborn1991hypergeom}). Definition \ref{def:GL} need not be restricted to real values of $\alpha$, and is valid for orders of differintegration $\alpha \in \mathbb{C}$. However, we restrict our attention at present to real orders. Furthermore, in general practice, $\alpha$ is most often restricted to the interval $(-1,2)$, as in \cite{mathieu2003fractional,baleanu2010fractional,elaraby2016fractional}.

The numerical algorithm simply ignores the limit as $h \to 0$, giving us the first GL algorithm. 

\begin{equation}\label{eq:GLnumerical}
	\left[ D^\alpha f(x) \right]_{GL} = h^{-\alpha} \sum_{k=0}^{N-1} b_kf(x-hk).
\end{equation}

As it stands, this algorithm is useful for differintegrating a function at the right endpoint of the function domain, $x$. However, the algorithm is adapted in order to differintegrate an entire array of function values. There are two attractive options for differintegrating arrays of function values: matrix methods and discrete convolutions.

\subsubsection{Matrix Method:}

The matrix method simply considers the task of differintegrating an array of function values by computing the differintegral at each function value, and saving each result as an element in a new array. To facilitate this, the array of function values may be expressed as a column vector and multiplied by a lower triangular differintegration matrix.

Let $f = [f(0),f(h), f(2h),\ldots, f(kh), \ldots, f(Nh)]^T$ be a vector of $N+1$ function values defined on a grid with step size $h$. The GL method may be applied to this vector by the following matrix multiplication.

\begin{equation}
	D^\alpha f(x) = h^{-\alpha} \cdot
	\begin{bmatrix}
    	b_0		&	0		&	0		&	0		&	0		&	\cdots	&	0 		\\
        0		&	b_0		&	0		&	0		&	0		&	\cdots	&	0 		\\
        0		&	b_1		&	b_0		&	0		&	0		&	\cdots	&	0		\\
        0		&	b_2		&	b_1		&	b_0		&	0		&	\cdots	&	0		\\
        0		&	b_3		&	b_2		&	b_1		&	b_0		&	\cdots	&	0		\\
        \vdots	&	\vdots	&	\vdots	&	\vdots	&	\vdots	&	\ddots	&	\vdots	\\
        0		&	b_{N-1}	&	b_{N-2}	&	b_{N-3}	&	b_{N-4}	&	\cdots	&	b_0	
    \end{bmatrix}
    \begin{bmatrix}
    	f(0) \\ f(h) \\ f(2h) \\ f(3h) \\ f(4h) \\ \vdots \\ f(Nh)
    \end{bmatrix},
\end{equation}

This method takes advantage of the efficient algorithms for matrix multiplication contained in the NumPy package \cite{walt2011numpy}. The computational complexity is the same as for multiplying an $(N+1) \times (N+1)$ lower triangular matrix by a vector of length $N+1$, and is $\mathcal{O}(N^2)$. For computational complexity, we follow the convention of including only floating point multiplications.

\subsubsection{Discrete Fourier Transform Method:}

In the problem of numerical differintegration, we define the discrete convolution filter with the entries as the $b_k$ coefficients from Definition \ref{def:GL}. Thus, the differintegration filter for the GL algorithm is given by

\begin{equation}
	D^\alpha = [b_0, b_1, \ldots, b_{N-2}, b_{N-1}],
\end{equation}

\noindent where $N+1$ is the number of function values. If $f$ is an array of function values, then the GL differintegration algorithm can be represented by the convolution

\begin{equation}
	\left[ D^\alpha f \right]_{GL} = h^{-\alpha}(f*D) = h^{-\alpha}\sum_i f_i \cdot D_{j-i},
\end{equation}

\noindent taken at each function value $f_j$, with $0 \leq j \leq N$. The convolution operation is identical in speed to the matrix multiplication method, $\mathcal{O}(N^2)$. Note that to perform the convolution, the coefficient filter is padded with zeros to ensure it is the same length as the function array.

For large enough $N$, it is possible to speed up the convolution operation by first computing the discrete Fast Fourier Transforms (DFTs) of the padded convolution filter and the function values, and then multiplying them together element-wise using the Fourier Transform property of the convolution, $F[A*B] = F[A]\cdot F[B]$. To obtain the result, the inverse DFT is then computed (see \cite{smith2007dft}). Let $F[\cdot]$ represent the DFT of the function $f$ and $F^{-1}[\cdot]$ represent its inverse DFT. Thus, the GL algorithm can be represented as 

\begin{equation}
	\left[ D^\alpha \right]_{GL} = h^{-\alpha}\left( F^{-1}[F[f]\cdot F[D]] \right).
\end{equation}

The method of differintegrating with the DFT convolution has complexity of order $\mathcal{O}(N \log N)$, which represents a significant computational speed-up for large $N$. If $N$ is not a multiple of 2, both the convolution filter and the function array can be padded with zeros to fully take advantage of this improvement in speed.

\subsection{The GLI Algorithm}
To improve the convergence properties of the GL definition, the following definition was proposed in \cite{oldham1974fractional}.

\begin{definition}
	Let $f(x)$ and its domain be defined as in Definition \ref{def:GL}. The improved Gr\"unwald-Letnikov (GLI) differintegral of order $\alpha \in \mathbb{R}$ is defined as:
  \begin{equation}
      D^\alpha f(x) = \lim_{h\to 0} h^{-\alpha} \sum_{k=0}^{N-1} b_k f\left(x+\frac{\alpha h}{2} - kh\right).
  \end{equation}
\end{definition}

The GLI algorithm begins from the same point as the GL algorithm, ignoring the limit as $h \to 0$. However, the GLI algorithm uses 3-point Lagrange interpolation to approximate the function values $f\left(x+\frac{\alpha h}{2} - kh\right)$. To simplify the notation, let $f$ be an array of $N+1$ function values on the interval $[0,x]$, and let $f_j$ be the function value defined at the $j$\SP{th} grid point, $0 \leq j \leq N$. We will make use of the approximation from \cite{oldham1974fractional}, given by

\begin{equation}
	f_{j+\alpha/2} \approx f_j + \frac{\alpha}{4}(f_{j+1} - f_{j-1}) + \frac{\alpha^2}{8}(f_{j+1} - 2f_j + f_{j-1}).
\end{equation}

We thus obtain the GLI algorithm as follows.

\begin{equation}\label{eq:GLI_algorithm}
	\left[ D^\alpha f(x_j) \right]_{GLI} = h^{-\alpha} \sum_{k=0}^{j} b_k \left[ f_j + \frac{\alpha}{4}(f_{j+1} - f_{j-1}) + \frac{\alpha^2}{8}(f_{j+1} - 2f_j + f_{j-1}) \right]
\end{equation}

Note that this algorithm requires $N+2$ function values for the differintegral to be computed at the $N$\SP{th} data point. This is not a problem if we have an explicit expression for the function, as the additional data point can simply be calculated. However, if the data comes from real measurements, there are two options. First, the point can be extrapolated after fitting an approximating function, such as a Lagrange polynomial, to the data. Second, if the step size $h$ is very small, the algorithm can be stopped at the $N-1$\SP{th} data point, and the resulting numerical result can be used to estimate the value of the differintegral at the right endpoint.

By some factoring of the expression in \eqref{eq:GLI_algorithm}, we can see that there are three separate objects being used to compute the differintegral at the point $x_j$. First, we define an array of 3 coefficients 
\begin{equation}\label{eq:GLIinterpolat}
A = \left[\left( \frac{\alpha^2}{8} - \frac{\alpha}{4} \right), \left( 1 - \frac{\alpha^2}{4} \right), \left( \frac{\alpha}{4} + \frac{\alpha^2}{8} \right)\right],
\end{equation}
which represents the coefficients for the previous, current, and next function values, respectively ($f_{j-1},f_{j},f_{j+1}$). 

In practice, it is convenient to compute the GLI algorithm using the following steps.
\begin{itemize}
\item Compute the array of previous, current, and next coefficients $A$ as above.
\item Obtain the function array $F = [f_0, f_1, \ldots, f_j]$.
\item Obtain the coefficient array $B = [b_0, b_1, \ldots, b_{j-2},0,0]$.
\item Perform the convolution $F*B$.
\item Multiply element-wise $D = A \cdot (F*B)$.
\item Sum the elements of $D$ and multiply the result by $h^{-\alpha}$.
\end{itemize}

Using the above steps, the GLI algorithm has complexity of order $\mathcal{O}(N^2)$.

\subsection{The RL Algorithm}
The RL algorithm, stated in \cite{baleanu2010fractional}, uses a quadrature rule to approximate the value of the RL differintegral, defined as follows.

\begin{definition}
Let $f(x)$ and its domain be defined as in Definition \ref{def:GL}. The RL differintegral is defined as
\begin{equation}
  	D^\alpha f(x) = \frac{1}{\Gamma(-\alpha)} \int_0^x (x-t)^{-\alpha-1}f(t)dt,
\end{equation}
where $\Gamma(\cdot)$ is the gamma function.
\end{definition}

The numerical algorithm for the RL definition, as defined by Diethelm in \cite{diethelm1997algorithm}, uses piecewise linear interpolation to define the quadrature rule for the RL differintegral at the point $f(x_j)$ as follows.

\begin{equation}\label{eq:RLalgorithm}
	\left[ D^\alpha f(x_j) \right]_{RL} = h^{-\alpha}\sum_{k=0}^j A_{k,j} f(x_k),
\end{equation}

where the coefficients $A_{k,j}$ are given in \cite{baleanu2010fractional} as

\begin{equation}\label{eq:RLcoeff}
	A_{k,j} = \frac{1}{\Gamma(2-\alpha)}
    \begin{cases}
    	(j-1)^{1-\alpha} - (j+\alpha-1)k^{-\alpha}, & k = 0 \\
        (j-k+1)^{1-\alpha} + (j-k-1)^{1-\alpha} - 2(j-k)^{1-\alpha}, & 1 \leq k \leq j-1 \\
        1, & k=j.
    \end{cases}
\end{equation}

To code this algorithm, the $A_{k,j}$ coefficients are placed into an $N \times N$ lower triangular matrix shown in \eqref{eq:RLmatrix}. 

\begin{equation}\label{eq:RLmatrix}
	R = \frac{1}{\Gamma(2-\alpha)}\cdot
	\begin{bmatrix}
    	1		&	0		&	0		&	\cdots	&	0		\\
        A_{0,1}	&	1		&	0		&	\cdots	&	0		\\
        A_{0,2}	&	A_{1,2}	&	1		&	\cdots	&	0		\\
        \vdots	&	\vdots	&	\vdots	&	\ddots	&	\vdots	\\
        A_{0,N}	&	A_{1,N}	&	A_{2,N}	&	\cdots	&	1
    \end{bmatrix}
\end{equation}

The matrix-vector product is then computed with the matrix $R$ and the array of $N+1$ function values, $F$. This operation has computational complexity $\mathcal{O}(N^2)$. Thus, the RL algorithm can be condensed to the following matrix-vector product.

\begin{equation}\label{eq:RLalgorithmshort}
	\left[ D^\alpha f(x_j) \right]_{RL} = h^{-\alpha}R\cdot F
\end{equation}

\section{The \textit{differint} Package}
\label{sec:differint}

\begin{table}\centering
\begin{tabular}{ll}
\hline
Core function    & Usage \\
\hline
GLpoint      &  Computes the GL algorithm at a point. \\
GL          & Computes the GL algorithm for an array of function values,        \\
       & using the Fast Fourier Transform.      \\
GLI       & Computes the GLI algorithm for an array of function values.      \\
RLpoint &   Computes the GL algorithm at a point.    \\
RL          & Computes the RL algorithm for an array of function values. \\
\hline
\end{tabular}
\caption{Core functions included in the \textit{differint} package.}
\label{tb:dfcore}
\end{table}

\begin{table}\centering
\begin{tabular}{ll}
\hline
Auxiliary function    & Usage \\
\hline
Gamma       &   Computes the Gamma function to 15 decimal points. \\
isInteger      &  Determines if an input number is an integer. \\
checkValues          & Used for algorithm input type-checking.        \\
GLIinterpolat       & Class to define the interpolating coefficients in \eqref{eq:GLIinterpolat}.      \\
functionCheck       & Determines if algorithm function input is callable or an array,      \\
	&	defines function array, and sets step size.	\\
poch          & Computes the Pochhammer symbol. \\
GLcoeffs       & Defines the convolution filter for the GL algorithm.      \\
RLcoeffs &   Computes the RL coefficients $A_{k,j}$ from \eqref{eq:RLcoeff}.    \\
RLmatrix          & Defines the matrix $R$ as in \eqref{eq:RLmatrix}. \\
\hline
\end{tabular}
\caption{Auxiliary functions included in the \textit{differint} package.}
\label{tb:dfaux}
\end{table}

This package is used for numerically computing the algorithms described in section~\ref{sec:algorithmdescription}. Through the API, the user can compute the various differintegrals at a point or over an array of function values. At present, there is little in the way of readily available, easy-to-use code for numerical fractional calculus. What is currently available are functions that are generally either smart parts of a much larger package, or only offer one numerical algorithm. The \textit{differint} package offers a variety of algorithms for computing differintegrals and several auxiliary functions relating to generalized binomial coefficients.

The full list of core functions is included in Table \ref{tb:dfcore}, and the list of auxiliary functions is included in Table \ref{tb:dfaux}.

\subsection{Examples}
\label{sec:examples}

To test the accuracy of the algorithms in the \textit{differint} package, three elementary functions were differentiated with order $\alpha = 1/2$/. To simplify the hand calculations, all example functions were differentiated on the interval $[0,1]$, with 120 grid points. The results of the computations are presented in Table \ref{tb:results}.

The exact values of the $1/2$-order derivative for each of the candidate functions are listed as follows:

\begin{align}
	D^{1/2} (\sqrt{x}) |_{x=1} &= \frac{\sqrt{\pi}}{2} &\approx 0.886226925453 \\
    D^{1/2} (x^2 - 1) |_{x=1} &= \frac{5}{3\sqrt{\pi}} &\approx 0.94031597258 \\
    D^{1/2} (e^{x}) |_{x=1} &= e\cdot \text{erf}(1) + \frac{1}{\sqrt{\pi}} &\approx 2.85488783585
\end{align}

where erf($\cdot$) is the error function (see \cite[p.14]{seaborn1991hypergeom}).

\begin{table}\centering
\begin{tabular}{lllll}\hline
Function	&	Algorithm	&	Computed Value	&	Absolute Error	&	Relative Error	\\ \hline
			&	GL			&	0.889988356363	&	3.761431e-3	&	4.244320e-3	\\
$\sqrt{x}$	&	GLI			&	0.886308277526	&	8.135207e-5	&	9.179598e-5	\\
            &	RL			&	0.886317417031	&	9.049158e-5	&	1.021088e-4	\\ \hline
			&	GL			&	0.937708605184	&	2.607367e-3	&	2.772863e-3	\\
$x^2-1$		&	GLI			&	0.895915217344	&	4.440075e-2	&	4.721897e-2	\\
			&	RL			&	0.939961210942	&	3.547616e-4	&	3.772792e-4	\\ \hline
            &	GL			&	2.86263962909	&	7.751793e-3	&	2.715271e-3	\\
$e^{x}$		&	GLI			&	2.81390977101	&	4.097806e-2	&	1.435365e-2	\\
			&	RL			&	2.85441394392	&	4.738919e-4	&	1.659932e-4	\\ \hline
\end{tabular}
\caption{Numerical approximations of the $1/2$-derivative of various elementary functions on the interval $[0,1]$, with 120 grid points.}
\label{tb:results}
\end{table}

\section{Discussion and Conclusion}
\label{sec:conclusion}
For all of the elementary functions tested in section 3.1, the RL algorithm was consistently more accurate than both the GL and the GLI algorithms in terms of absolute error. The GLI algorithm suffered a loss in accuracy for the polynomial function and the exponential function. This loss in accuracy is due to the relatively large changes in function value over a small interval. Since the GLI algorithm is using the differintegral at the $N-1$\SP{th} function value to estimate the true value, this method is prone to exaggerated error for functions with large first derivatives. In the case of the square root function, the GLI algorithm performed as well as the RL algorithm and far better than the GL algorithm.

In this work, the \textit{differint} package was introduced. This package allows for the convenient computation of fractional derivatives and integrals in the Python programming language. Full source code and documentation for the \textit{differint} package can be found at https://github.com/differint/differint.

The code base for this project can be used for many applications, such as fractional edge detection, as in \cite{mathieu2003fractional,elaraby2016fractional}, and for computing the coefficients for series solutions to fractional differential equations as in \cite{baleanu2010fractional,diethelm1997algorithm}. Future iterations of this package will include the high-order Caputo approximation presented by Li, Cao, and Li \cite{li2016higherorder}.

\section{Acknowledgements}
The author expresses his sincerest thanks to Wenyuan Liao for his mentorship and direction, and to the University of Calgary for its support.

\bibliographystyle{sorted}
% \bibliography{main}



\end{document}